\documentclass{icrc2009}

\usepackage{graphicx}   
\usepackage[caption=false]{caption}    
\usepackage[font=footnotesize]{subfig} 
\usepackage{fixltx2e}
\usepackage{url}

\newcommand{\shorttitle}[1]%
{\markboth{Proceedings of the 31\MakeLowercase{$^{st}$} ICRC, {\L}\'{o}d\'{z} 2009}{#1} }
\newcommand{\etal}{\MakeLowercase{\textit{et al. }}} 

\begin{document}
\title{Upgrade of the VERITAS Cherenkov Telescope Array}

\author{\IEEEauthorblockN{A.~Nepomuk Otte\IEEEauthorrefmark{1}\\for the VERITAS collaboration\IEEEauthorrefmark{2}}\\
\IEEEauthorblockA{\IEEEauthorrefmark{1}Santa Cruz Institute for Particle Physics, University of California at Santa Cruz, CA 95060 Santa Cruz, U.S.A.\\ (nepomuk@scipp.ucsc.edu)\\
\IEEEauthorrefmark{2}see R.~A.~Ong et al. (these proceedings)  or http://veritas.sao.arizona.edu/conferences/authors?icrc2009
 for a full author list
}

}
\shorttitle{Otte \etal VERITAS upgrade}
\maketitle

\begin{abstract}
 The VERITAS Cherenkov telescope array has been fully operational since Fall 2007 and has fulfilled or outperformed its design specifications. We are preparing an upgrade program with the goal to lower the energy threshold and improve the sensitivity of VERITAS at all accessible energies. In the \emph{baseline} program of the upgrade we will relocate one of the four telescopes, replace the photo-sensors by higher efficiency photomultipliers and install a new trigger system. In the \emph{enhanced} program of the upgrade we foresee, in addition, the construction of a fifth telescope and installation of an active mirror alignment system.
  \end{abstract}


\section{Introduction}

Gamma-ray astronomy with air-shower imaging Cherenkov telescopes (IACTs) is a young discipline. With the latest generation of instruments much insight has been gained into the non-thermal processes in a variety of astrophysical sources, such as, for example, active galactic nuclei, supernova remnants, and  pulsar wind nebulae. Observations in gamma-rays above 100 \,GeV also provide a tool to study interesting questions that are of fundamental importance in physics and cosmology. Some examples are the nature of dark matter and the density and evolution of the extragalactic background light.

\begin{figure}[!htb]
 \centering
 \includegraphics[width=3.2in]{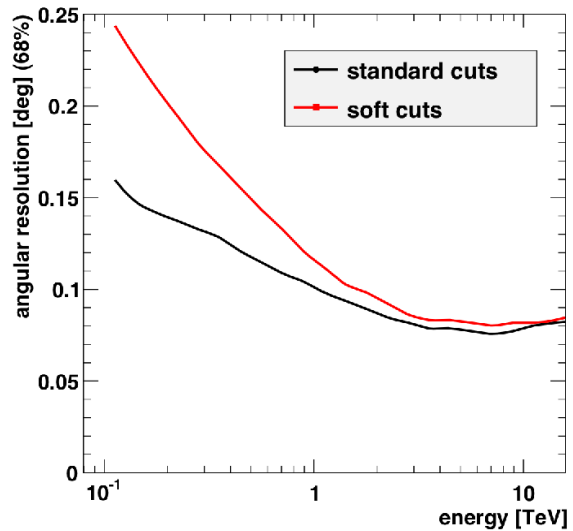}
  \caption{Angular resolution as function of energy. "standard" cuts refer to gamma-ray selection cuts suitable for sources with an energy spectrum similar to or harder than the Crab Nebula. ``soft" cuts are used for sources with correspondingly softer spectra.}
 \label{AngRes}
 \end{figure}

Of the existing IACTs the VERITAS Cherenkov telescope array is one of the most sensitive gamma-ray instrument above 100\,GeV. VERITAS can detect a point source with a power law spectrum of index -2.5 and 1\% of the Crab Nebula flux in less than 50 hours at a significance level of $5\sigma$. The angular and energy resolutions are energy dependent. At 1 TeV the reconstructed arrival direction for 68\% of all gamma-ray events fall within $\sim0.1^\circ$ of the true arrival direction (cf.~Figure \ref{AngRes}); the energy resolution at this energy is $\sim15\%$.

The VERITAS array has achieved, and in some metrics surpassed, its design specifications and science goals; however, the performance parameters are not limited by any fundamental physical constraints. Improved performance can be achieved through improved data analysis
techniques and hardware upgrades, which will allow us to better address the key science goals.

In this paper we describe a staged upgrade program for VERITAS, to envisaged over the next few years, with the primary goals to lower the energy threshold of the instrument and to increase its sensitivity; both of these will allow us to better address our key science questions.

We describe a \emph{baseline} program, which includes:

 \begin{itemize}
 \item relocating one of the telescopes,
 \item refitting the telescope cameras with more sensitive photo-detectors, and
 \item implementing a different hardware trigger scheme.
 \end{itemize}

A more ambitious \emph{enhanced} program is also described which, in addition to the baseline upgrades, involves the construction of an additional telescope and the development of an active mirror alignment system.

\section{Baseline Program}

The \emph{baseline} program includes the relocation of one telescope, an upgrade of the focal plane instrumentation with high efficiency photon detectors, and a new trigger system. The first step, the relocation of telescope T1 is presently ongoing.

\subsection{Relocation of Telescope T1}

 \begin{figure}[!htb]
 \centering
 \includegraphics[width=3.2in]{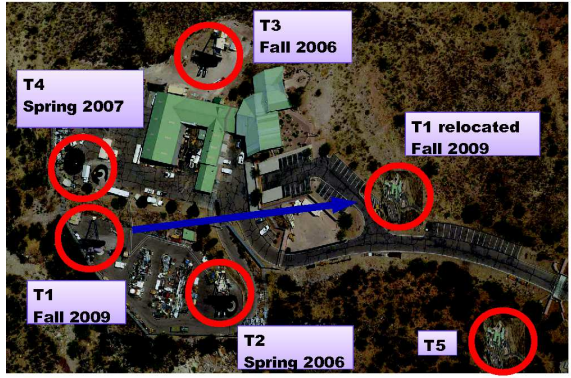}
  \caption{A plan view of the Whipple Observatory Basecamp, taken during the construction of telescope 4. The four existing VERITAS telescope positions are indicated, along with the planned position of telescope 1 after relocation and a potential site for a new telescope, T5.}
 \label{VERITAS_array}
 \end{figure}

The VERITAS array is located at the basecamp of the Whipple Observatory in southern Arizona. The first telescope was placed there in 2002-2003 with the anticipation to relocate the array to a different location. It was subsequently decided to build the whole array at the same site, which resulted in the current layout that is shown in Figure \ref{VERITAS_array}. With a baseline of 35\,m, T1 and T4 allowed some interesting studies of the effect that the close proximity of two Cherenkov telescopes has on the sensitivity of the array around 100\,GeV \cite{Kildea08}. At the same time, the proximity of T1 and T4 results in a suboptimal effective collection area at higher energies, which can be maximized by a more uniformly shaped rectangular configuration of the array.

\begin{figure}[!htb]
 \centering
 \includegraphics[width=3.2in]{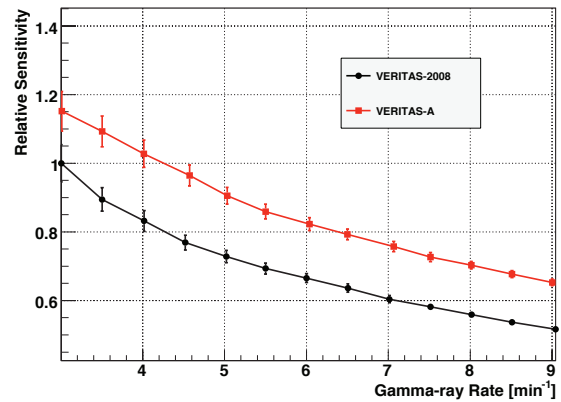}
  \caption{Effect of the relocation of Telescope 1 on the sensitivity as a function of the gamma-ray rate after cuts, normalized to the peak sensitivity for the current layout of the array. ``VERITAS-2008"  and  ``VERITAS-A" refer to the configuration before and after the relocation, respectively.}
 \label{RelocImprov}
 \end{figure}

Detailed simulations have shown that the sensitivity of the array increases by 15-20\% (cf.~Figure \ref{RelocImprov}) if telescope T1 is relocated to the position indicated in Figure \ref{VERITAS_array}. The most important practical result of this relocation is that the time required to detect a weak source is reduced by about 30\%, which is equivalent to extending the possible observation time from an annual 1000 hours to 1400 hours annually.

The relocation of T1 is presently under way and expected to be finished by the beginning of the upcoming observation cycle in fall 2009.

\subsection{Higher Efficiency Photon Detectors}

We have considered several possible upgrades for the VERITAS telescope cameras:
\begin{itemize}
\item an extension of the field of view,
\item a smaller pixellation, and
\item improved photon collection efficiency.
\end{itemize}
These are all desirable improvements but the most effective, practical, and economical upgrade is to increase the photon collection efficiency by replacing the existing PMTs with higher quantum efficiency devices that have become commercially available. These PMTs have peak photon detection efficiencies of about 35\% in the blue, as compared to a peak efficiency of 18\% of the devices currently in use on VERITAS.

\begin{figure}[!htb]
 \centering
 \includegraphics[width=3.0in]{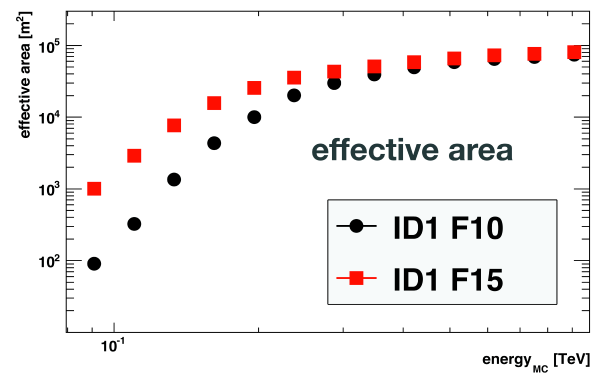}
  \caption{Effective area of VERITAS after standard analysis; dots for the current configuration and squares for the expected improvement after an upgrade with high efficiency PMTs.}
 \label{HighQEImprov}
 \end{figure}

Replacing the photon detectors with these higher efficiency ``super bialkali" photomultipliers would result in an increase of the optical throughput for Cherenkov light from an air shower by about 40\%. Our simulations show that the immediate effect of this upgrade would result in about 17\% improvement in the sensitivity of the array for a weak point source above 100\,GeV. Of possibly higher importance is the increase of sensitivity of the array at and below 100\,GeV where the  effective area is expected to increase by about one order of magnitude if our standard analysis tools are applied (see Figure \ref{HighQEImprov}). If an analysis tuned for higher sensitivity at 100\,GeV is used we expect an overall significant increase of the effective area at 100\,GeV.

\subsection{A Topological Trigger}

Additional improvement of the performance at lower energies is expected by implementing a more intelligent hardware trigger system as part of the baseline VERITAS upgrade program. The trigger threshold in Cherenkov telescopes is largely dependent on the ability to suppress accidental triggers caused by fluctuations in the night-sky background light (NSB) and triggers caused by cosmic rays, mostly protons, muons, and electrons. While stereoscopic operation already provides excellent suppression of background because of tight trigger coincidence windows of 100\,ns or less between telescopes, we expect further improvements as the result of two parallel trigger system research and development efforts that have been going on over the past few years.

In one effort we developed an FPGA based simple three-fold neighbor logic that achieved about 4\,ns resolving times in a test on one of the telescopes where the system was running for one month in 2008. A 4\,ns coincidence window would allow us to reduce the  trigger rate caused by fluctuations in the NSB by more than a factor of two.

 In another effort, we are following a different approach to develop a three-stage, high-speed trigger to suppress accidentals from both the NSB and cosmic rays. This ``topological trigger" \cite{martin,toptrigger} consists of high-speed camera-level triggers that transmit image parameters to the array trigger for real-time event classification. The topological trigger is a shift from off-line to real-time gamma/hadron separation.

\section{Enhanced Program}

\subsection{Array Expansion}

An enhanced upgrade option would be the construction of a  fifth telescope, with the obvious impact of an increase in the effective area. This increase would in turn improve the sensitivity by an amount approximately proportional to the square root of the number of telescopes in the array.

In reality, an improvement in sensitivity larger than 12\% is expected because of the better angular resolution \cite{HintonMC} and better gamma-hadron separation. A better angular resolution will allow for better suppression of background events and will have a bigger impact  on morphological studies of Galactic sources and the study of possible gamma-ray of halos around extragalactic sources. Another advantage of a five-telescope array would be the operation in split mode, e.g.~the parallel observation of two sources with two and three telescopes, respectively.

Detailed studies with Monte Carlo simulations are under way to investigate the various aspects and impacts that the addition of a fifth telescope would have.

\subsection{Active Mirror Alignment}

     Each VERITAS telescope has 350 mirror facets. A realignment becomes necessary every time a facet is recoated, which is done approximately every two years.
    The alignment procedure is done manually \cite{andrew}, which is manpower intensive and better done in an automated way. An automated alignment procedure would also correct for elevation-dependent mechanical deformations of the optical support structure. This task can be accomplished by an active mirror alignment system (AMAS).

An AMAS would  eliminate the aforementioned disadvantages of the alignment method currently used and, in addition, would allow monitoring the reflectivity of individual mirror facets \cite{MAGIC_ICRC}, changing the focus from infinity to the height of the shower maximum, and altering the optical axis of the OSS to compensate systematic effects, for example due to camera sagging. Finally, the ability to defocus the OSS provides an additional safety feature in the case of a telescope drive failure.

\section{Discussion}

The VERITAS gamma-ray observatory in its present configuration has achieved its design specifications and is one of the most sensitive gamma-ray instruments. In this paper we have outlined an upgrade program, which would allow us to increase the sensitivity and lower the energy threshold of VERITAS.

 In summary the upgrade program comprises:
\begin{itemize}
\item relocating one of the VERITAS telescopes,
\item upgrading the VERITAS cameras with photon detectors of higher efficiencies, and
\item implementing a novel trigger system.
\end{itemize}
The enhanced upgrade program includes all of the above options plus
\begin{itemize}
\item adding a telescope, and
\item automatic mirror alignment system.
\end{itemize}

The discussed upgrade options will result in a significant improvements of the performance of VERITAS and ultimately on all science topics that we address with VERITAS. A lower energy threshold will allow us to look deeper into the Universe, and at the same time increases the overlap in energy with the gamma-ray satellite Fermi.

A higher sensitivity at all energies will provide us with the possibility to observe more sources or perform deeper observations of particular objects. The improved sensitivity after the relocation of telescope T1 is in part due to a better angular resolution, which will also allow us to make morphological studies with higher resolutions than presently possible.

Of all the discussed upgrade options the move of T1 has been fully funded and is presently ongoing. All other upgrade options are still under discussion with parallel efforts in simulations and research and development.

\section{Acknowledgment}

This research is supported by grants from the US Department of
Energy, the US National Science Foundation, and the Smithsonian
Institution, by NSERC in Canada, by Science Foundation Ireland, and
by STFC in the UK. We acknowledge the excellent work of the technical
support staff at the FLWO and the collaborating institutions in the
construction and operation of the instrument.

\end{document}